# Tailoring of multi-pulse dynamics in mode-locked laser via optoacoustic manipulation of quasi-continuous-wave background

Ki Sang Lee, Chang Kyun Ha, Kyoung Jun Moon, Dae Seok Han, and Myeong Soo Kang[*]

*Department of Physics, Korea Advanced Institute of Science and Technology (KAIST)*

*291 Daehak-ro, Yuseong-gu, Daejeon 34141, Republic of Korea*

[*]*mskang@kaist.ac.kr*

A variety of nonequilibrium multi-pulse states can emerge in a mode-locked laser through the interactions between the quasi-continuous-wave background (qCWB) and optical pulses inside the laser cavity. However, they have been long regarded as unpredictable and hardly controllable due to the noise-like nature of the qCWB, and relevant previous studies thus lack a clear understanding of their underlying mechanisms. Here, we demonstrate that the qCWB landscape can be manipulated via optoacoustically mediated interactions between the qCWB and mode-locked pulses, which dramatically alters the behaviors of multi-pulse dynamics in unprecedented manners. In this process, impulsive qCWB modulations are created at well-defined temporal locations, which act as the point emitters and attractive potentials for drifting pulse bunches and soliton rains. Hence, we can transport a single pulse bunch from a certain temporal position to another on the qCWB, and also make the soliton rain created and collided exclusively at specific temporal locations, in sharp contrast to the conventional cases. Our study opens up new possibilities to control the nonequilibrium multi-pulse phenomena precisely in the time domain, which would not only help the observation and clear understanding of undiscovered features of multi-pulse dynamics but offer a practical means of advanced optical information processing.

# INTRODUCTION

Multi-soliton dynamics has attracted great attention in a broad range of scientific disciplines, e.g., classical and quantum fluid dynamics, plasma physics, magnetism, complex networks, chemical reactions, neurosciences, and nonlinear optics [1]. Precise control of the behaviors of multiple solitons in the temporal or spatial domain is highly beneficial, because it offers not only an essential experimental tool for deeper understanding of multi-soliton interactions and resulting many-body nonlinear phenomena but also a practical means of advanced information processing. In the area of nonlinear optics, mode-locked fiber lasers have been widely utilized not only as versatile sources of ultrashort optical pulses but also as fertile grounds for exploring a variety of nonlinear multi-pulse dynamics (MPD) [1,2]. In these systems, the interplay among the optical nonlinearities, dispersion and spectral filtering, and the inherently dissipative nature of laser cavities with the mutually counteracting gain and loss allow us to access rich MPD such as harmonic mode locking [3,4], bound state formation [5,6], pulse bunching [7,8], and soliton collisions [9].

An important nonlinear optical effect that mediates the MPD is the long-range optoacoustic interactions among the intracavity optical pulses through the optically driven transverse acoustic resonances (ARs) in the fiber [10]. As a mode-locked optical pulse circulates inside the single-mode fiber laser cavity, it excites electrostrictively a sequence of AR impulses in the fiber, their temporal widths and separations being typically a few nanoseconds and a couple of tens of nanoseconds, respectively [11]. They create a series of impulsively time-varying refractive index modulation, which gives rise to the spectral shifts of the subsequent optical pulses. Their group velocities then change via dispersion, which yields the temporal shifts of the pulses [12–14]. Furthermore, the optoacoustic interactions in standard single-mode fiber are significant over the AR frequency range of several tens to hundreds of MHz, which in some cases acts as a limiting factor of the repetition rate of harmonic mode-locking up to hundreds of MHz [3,4,15]. It has been recently demonstrated that the use of gigahertz ARs tightly confined in the micron-sized core of photonic crystal fiber can solve this issue, stabilizing the harmonically mode-locked pulse train at the repetition rate beyond 1 GHz [16–18]. The formation of pulse bunches and bound pulse states via the optoacoustic inter-pulse coupling has also been investigated [19–22].

Another crucial mechanism that significantly influences the MPD is the noise-like fluctuation of the quasi-continuous-wave background (qCWB) that spreads over the entire laser cavity [23–28]. The qCWB can be developed from the random beats of longitudinal cavity modes [23] or the spontaneous emission from the laser gain medium [29]. The qCWB fluctuation, in general, gives rise to the diffusive motion (or timing jitter) of the co-propagating mode-locked pulses [24], which yields the long-range interactions among the pulses in the presence of slow gain dynamics [25]. Soliton rain (SR) is a representative example of the resulting nonequilibrium multi-pulse states, where soliton pulses are created spontaneously from the fluctuating qCWB and drift toward a pre-existing soliton bunch [23]. The qCWB fluctuation can also reportedly mediate the damped oscillation of pulse parameters and the pulse bunch formation via the long-range attractive inter-pulse interactions [24,25]. The effects of the qCWB have also been investigated for the cases in which continuous-wave laser light externally injected into the laser cavity plays a role of the qCWB [26–28]. The qCWB can then interfere with the dispersive waves shed from the soliton pulses, which leads to the drift motion of the soliton pulses. Such the nonequilibrium multi-pulse phenomena, however, have been long regarded as unpredictable and hardly controllable due to their stochastic natures, and relevant previous studies thus lack a clear understanding of their origins and underlying mechanisms.

In this paper, we address these problems by considering an intriguing situation where the inter-pulse optoacoustic interactions and the qCWB fluctuations exert a considerable impact simultaneously on the MPD in a correlated fashion. Such the possibility has not been studied yet, and there is thus still a completely unexplored regime in which the optoacoustic interactions might critically govern the behaviors of MPD. We experimentally show that the qCWB landscape can be manipulated via the optoacoustically excited ARs in the single-mode fiber laser cavity, which in turn dramatically alters the nonequilibrium MPD. In this process, a set of impulsive qCWB modulations are created in the form of intensity peaks or dips separated by the characteristic temporal periods of ARs (e.g., 21 ns and 33 ns for the 125-μm-thick fiber cladding), which we can adjust potentially by changing the thickness of AR-trapping optical fiber. In particular, we demonstrate that each qCWB modulation then acts as a point emitter and an attractive potential for the drifting pulse bunches and the SR, which provides a novel means to control the nonequilibrium MPD precisely in the time domain in unprecedented manners. We present a physical explanation on the formation of the optoacoustic qCWB fluctuations and describe how they play a critical role in the emergence of the diverse behaviors of nonequilibrium MPD in mode-locked fiber lasers.

# RESULTS

## Optoacoustically created impulsive qCWB modulation

We construct a mode-locked fiber laser with unidirectional ring cavity configuration, as shown in Fig. 1, where the passive mode locking is achieved based on nonlinear polarization rotation (NPR). A 78-cm-long section of erbium-doped fiber (EDF, absorption coefficient: 110 dB/m at 1530 nm) is used as a gain medium, which is pumped by a 976 nm laser diode through a wavelength division multiplexer. The total cavity length and net cavity dispersion are controlled simultaneously by adjusting the lengths of standard single-mode fiber (SMF, $\beta_2 = -22$ ps$^2$/km at 1550 nm) and dispersion compensating fiber (DCF, $\beta_2 = +128$ ps$^2$/km at 1550 nm) individually. The total cavity length is in the range of 30–40 m, which corresponds to the cavity round-trip time of 150–200 ns. The laser output is obtained via a 10% tapping coupler and then observed with a 12 GHz-bandwidth photodetector and a 2.5 GHz-bandwidth oscilloscope.

The threshold pump power for mode locking varies in the range of 40–70 mW depending on the net cavity dispersion $\beta_C$. Above the threshold, we can generate a mode-locked pulse train and a co-propagating qCWB simultaneously, as shown in Fig. 2(a), by carefully adjusting the intracavity polarization controllers (PCs). This state is readily produced at $\beta_C < -2$ ps$^2$/km, i.e., unless the cavity dispersion is normal or too weakly anomalous. The laser then operates at the soliton regime, and the intracavity pulse energy is set typically as 0.1–0.2 nJ. The existence of the qCWB is further verified by the relatively narrow spectral peak at ~1560 nm on top of the broad pulse spectrum that in some cases exhibits the Kelly sidebands (Fig. 2(b)) [30] or the spectral modulation originating from the quasi-periodic internal structure of the mode-locked pulse bunch [31]. We note that at high pump powers well above the mode locking threshold the laser can emit a train of pulse bunches [8,32] at the fundamental repetition rate. In the oscilloscope traces, however, each bunch is not fully resolved and seen as a single pulse because of the insufficient bandwidth of the oscilloscope. Nevertheless, we can determine the temporal spacing between the adjacent pulses in a bunch from the period of the spectral modulation in the optical pulse spectrum [31], which is typically a few picoseconds in our case. Furthermore, we can estimate the total number of pulses inside each bunch from the laser output power, which is generally on the order of 10 depending on the pump power and the qCWB level and can go beyond 100 when the pump power is above 300 mW while the qCWB is suppressed.

Focusing on the qCWB part (green dashed box in Fig. 2(a)), we observe sharp intensity modulation of the qCWB, as shown in Fig. 2(c). The qCWB modulation peaks at almost regularly spaced temporal locations, and their spacing measured as 33 ns or 21 ns is interestingly close to the temporal period of the previously reported optoacoustic impulse response to a single optical pulse via electrostriction in a 125-μm-thick optical fiber [11]. These peaks are a part of the qCWB rather than that of the mode-locked pulse train, which is seen from the fact that they can be observed only when the qCWB exists. The most marked group of qCWB peaks that appear integer multiples of $\tau_{TR}$ = 33 ns after a mode-locked pulse (green triangles in Figs. 2(c)–2(e)) can be produced by the torsional-radial $TR_{2m}$ ARs only. On the other hand, another group of the relatively weak qCWB peaks appearing at integer multiples of 21 ns (blue triangles in Figs. 2(c)–2(e)) can be contributed by both the $TR_{2m}$ and the radial $R_{0m}$ ARs. There are also qCWB peaks at 16 ns, 49 ns, 82 ns, and 116 ns from the mode-locked pulse (green ellipses, most apparent in Fig. 2(c)), which originate from the optoacoustic response to the mode-locked pulse in the previous round-trip. From the cavity round-trip time of 151 ns, their actual temporal positions are determined as 167 ns, 200 ns, 233 ns, and 267 ns, respectively, which belong to the 33 ns $TR_{2m}$ branch. Such the sharp qCWB modulations can be observed over wide ranges of cavity parameters such as the cavity length, cavity dispersion, and pump power, with careful adjustment of the intracavity PCs. We check that the peak intervals of 33 ns and 21 ns are almost unaffected by the change of all these cavity parameters, which further confirms that they are relevant to the AR properties rather than the optical ones, although the detailed pattern of the qCWB modulation may vary. In addition, we do not see any apparent correlation between the optical spectrum of the laser output and the features of the qCWB modulation in the time domain. We note that we can readily observe the dominant qCWB peaks at integer multiples of $\tau_{TR}$ = 33 ns for $\beta_C$ < -2 ps$^2$/km. In particular, at the relatively weak anomalous regime of -6 ps$^2$/km < $\beta_C$ < -2 ps$^2$/km, only the 33 ns branch of qCWB peaks is predominantly observed, whereas the 21 ns branch is rarely seen. Although the 21 ns branch becomes apparent at the cavity dispersion more anomalous beyond -6 ps$^2$/km, we do not investigate it in detail, as it does not affect the MPD significantly in our case. We also investigate the pure qCWB state, which we generate by suppressing the mode locking through the control of the intracavity PCs. The electrical spectrum of the pure qCWB state also exhibits the features of the $TR_{2m}$ ARs only rather than the $R_{0m}$ ones, which we describe in more detail in Supplementary Information.

One might think that the AR-induced phase modulation [33–36] could give rise to the qCWB peaks through interferences. The 21 ns $R_{0m}$ branch of qCWB fluctuations should be then predominant compared to the 33 ns one, as the magnitude of the optoacoustic impulse response mediated by the $R_{0m}$ ARs is stronger than that by the $TR_{2m}$ ones approximately by a factor of 10 or larger [11]. In our case, however, the 33 ns $TR_{2m}$ branch is stronger and appears more readily, whereas the 21 ns one is relatively weak and observed intermittently. This strong contradiction rejects the possibility of the AR-induced phase modulation as the primary origin of the qCWB peaks. On the other hand, it is known that the $TR_{2m}$ ARs can also yield the polarization modulation through depolarized scattering [10]. In our NPR-based mode-locked fiber laser, the intracavity polarizer can transform the polarization modulation into the intensity modulation, which results in the qCWB modulation.

To verify and further understand the role of the TR-AR-induced polarization coupling and intracavity polarizer in the formation of the qCWB modulation, we construct another mode-locked fiber laser that is very similar to Fig. 1 but incorporates a semiconductor saturable absorber mirror (SESAM) of as a mode locker and excludes the polarizer. The SESAM has a relaxation time of 2 ps, and its small-signal absorbance and modulation depth are 12% and 7%, respectively, which are appropriate for the simultaneous generation of the mode-locked pulses and qCWB. We can reproduce a set of impulsive qCWB fluctuations in this case. In contrast to the case of the NPR-based mode locking, however, it turns out that they appear mostly at integer multiples of $\tau_R$ = 21 ns at any cavity parameters, whereas their creation at integer multiples of $\tau_{TR}$ = 33 ns is hardly observed, as shown in Fig. 3(a). This result points out that the intracavity polarizer plays a critical role in the formation of the optoacoustic qCWB fluctuations spaced by $\tau_{TR}$ = 33 ns. For further comparison to figure out the role of the intracavity polarizer, we also generate a continuous-wave output from the laser after excluding the intracavity polarizer and the SESAM to suppress any type of passive mode locking and measure the electrical spectrum of the laser output. Unlike the electrical spectrum of Fig. S1(b) in Supplementary Information, we observe a group of spectral peaks at the integer multiples of either 30 MHz (Fig. 3(b)) or 47 MHz (Fig. 3(c)), depending on the state of intracavity PCs and the pump power. These two characteristic peak-to-peak separations can also appear simultaneously (Fig. 3(d)). The frequency spacings of 47 MHz and 30 MHz correspond to the modulation periods of 21 ns and 33 ns, respectively, which indicates that in the absence of the intracavity polarizer both the radial and the torsional branches can be revealed in the qCWB.

**Tailoring of MPD via optoacoustically manipulated qCWB**

We observe various types of novel MPD phenomena that are mediated by the TR-AR-induced optoacoustic qCWB fluctuations. As we will discuss later, the qCWB peaks can act as attractive potentials for optical pulses in the anomalous dispersion regime. First, it is then possible to trap a single pulse (or pulse bunch) at one qCWB peak and transport it to another, as described in Figs. 4(a)–4(c). After generating a train of mode-locked pulses and the qCWB simultaneously, we can produce additionally a relatively small pulse at one optoacoustic qCWB peak (located at $3\tau_{TR} = 100$ ns from the main mode-locked pulse in the case of Fig. 4(a)), by carefully adjusting the intracavity PCs. We can then perturb the trapped small pulse to release it from the qCWB peak and drift, as shown in Fig. 4(b), by slightly changing the state of the intracavity PCs. We experimentally observe that the drift speed tends to get smaller as the magnitude of the net cavity dispersion $|\beta_C|$ decreases. We can understand this tendency from the fact that the drifting pulse has a slightly different center wavelength from the main mode-locked pulses [38], and a larger value of $|\beta_C|$ yields a greater difference of group velocity between them, i.e., a higher drift speed. When the drift is sufficiently slow, the small pulse can be captured at the next optoacoustic qCWB peak (located at $4\tau_{TR} = 133$ ns in the case of Fig. 4(c)). These observations indicate a new possibility that we can make use of the optoacoustic qCWB fluctuations to control the motion of small pulses on the qCWB landscape precisely in the time domain, in the way of storing and transferring them at specific temporal positions that we target.

We can also generate several small drifting pulses that are distributed initially over the entire qCWB landscape, as shown in Fig. 4(d), by adjusting the intracavity PCs at the pump powers well above the mode-locking threshold. Although the pulses drift seemingly in a random way at the beginning, their motions eventually stop at the series of qCWB peaks or the main pulses to yield a regularly spaced pulse pattern, as shown in two examples of steady states of Figs. 4(e) and 4(f). This type of pulse pattern formation is observed more readily when the drift of small pulses is sufficiently slow, and the optoacoustic qCWB peaks are more apparent. This condition is achieved in our case when the net cavity dispersion $\beta_C$ is slightly anomalous but not too small so as to keep the simultaneous generation of the main mode-locked pulses and the qCWB, typically within the range of -7 to -2 ps$^2$/km. We note that in the presence of the qCWB dips that mostly appear at the relatively more anomalous regime (i.e., at larger values of $|\beta_C|$), not to be presented here in detail, we usually observe the chaotic drift of pulses that are created spontaneously from the qCWB in an uncontrollable manner.

The formation of the attractive potentials at which the drifting pulses are eventually captured can be contributed in general by two mechanisms. First, the optical pulses nearby the optoacoustically driven acoustic impulses experience frequency shifts through the sharp photoelastic index fluctuations, which then act as the attractive potentials under the anomalous dispersion. While this type of long-range pulse-to-pulse interactions has been recently investigated in the context of temporal cavity solitons [14,20], the previous works considered only the radial $R_{0m}$ ARs with the fully scalar analysis. On the other hand, the $TR_{2m}$ ARs that play a predominant role in our case yield much weaker photoelastic index changes by a factor of 1/10 or less compared to the radial ARs [11]. Hence, if the photoelastic frequency shift created by the acoustic impulses is the dominant mechanism of the pulse trapping, it would take place much more readily at the temporal locations of the integer multiples of 21 ns that is the characteristic of the R0m ARs rather than of the integer multiples of 33 ns. Another possible physical mechanism is the interaction between the drifting pulses and the intensity peaks on the qCWB via the cross-phase modulation (XPM). The drifting pulses experience larger refractive indices at the qCWB peaks through the XPM compared to outside of the qCWB peaks, so the pulses can be trapped at the qCWB peaks. We can estimate the frequency shift, i.e., the temporal rate of phase shift, arising from each of the two mechanisms for comparison. In our experimental conditions at the nominal pump power of 300 mW, the power modulation at the optoacoustic qCWB peak is measured as typically ~30 mW (Fig. 4(b)). The XPM-induced phase shift at the qCWB peak per unit propagation length in our cavity (effective cavity nonlinearity: $\gamma_{eff}$ = 2.0 $W^{-1}km^{-1}$) is then 0.12 rad/km. On the other hand, the AR-induced photoelastic phase shift per unit propagation length is estimated below 0.08 rad/km for the typical pulse energy of 0.9 nJ of each mode-locked pulse bunch [39,40]. Both types of phase shifts take place within a similar time scale of ~1 ns, so in our case the XPM-induced pulse frequency change is larger than or comparable to the photoelastic counterpart.

We also observe another type of interesting MPD phenomena, where a variety of novel SR dynamics emerges as the TR-AR-induced optoacoustic qCWB fluctuations radically alter the SR. The SR is formed when the pulse-like fluctuations created on the qCWB from the random beats of longitudinal laser modes are developed into soliton pulses through instantaneously saturated losses [23]. Hence, the SR can be produced in principle at almost arbitrary temporal locations on the qCWB landscape, and the resulting drifting pulses are absorbed eventually at the main mode-locked pulses. In strong contrast to the conventional SR, in our case the optoacoustic qCWB peaks act as the discrete 'point sources' of the SR, i.e., the SR is emitted from the qCWB peaks only, whereas its generation from the other temporal positions is highly suppressed, as shown in Figs. 5(a) and 5(b). Furthermore, we can make the SRs produced simultaneously at multiple qCWB peaks, as shown in Fig. 5(c). The optoacoustic qCWB peaks boost the loss saturation that develops the pulse-like fluctuations to the solitons. The SR is then much likely to be generated exclusively at the qCWB peaks. While the SRs have been regarded as irregular and hardly controllable in most cases, our observations suggest that we can engineer the SR emission by optoacoustically manipulating the qCWB landscape. To the best of our knowledge, this is the first-time experimental demonstration of control of the SRs by use of an intracavity source, rather than externally injected continuous-wave beams [23] or noise sources [41].

Finally, in a similar fashion to the capturing of a single drifting pulse in Figs. 4(a)–4(c), the qCWB peaks can also catch and hold the SR. As shown in Fig. 5(d), the SR emitted from a certain qCWB peak can be captured at the neighboring qCWB peak, where a secondary pulse bunch develops as the energy of the SR is accumulated there. This behavior of SR is more readily observed at the net cavity dispersions in the range of -7 to -2 $ps^2$/km, when their drift speed is small enough and the qCWB peaks are more apparent, as the case of trapping of a single drifting pulse in Figs. 4(a)–4(c). When the pulse bunch grows too high as the SR is continuously absorbed, it is forcedly released and drifts toward the main mode-locked pulse bunch, and the overall process is repeated, a new secondary pulse bunch growing again at the same qCWB peak. On the other hand, when the drift speed of the SR is not sufficiently low, which is more likely to take place at larger values of $|\beta_C|$, the SR can be hardly trapped at the qCWB peak. Instead, the drift speed is temporarily reduced and recovered as the SR passes through the qCWB peak, and the SR collides eventually at the main mode-locked pulse bunch. Our observations suggest the existence of the internal motions of the SR that have not been discovered yet, probably mediated by the ARs, besides the internal pulse bunching in the SR regime [22].

## CONCLUSION

We have reported the first-time experimental demonstration of tailoring of nonequilibrium multi-pulse phenomena in a mode-locked fiber laser through optoacoustic manipulation of the qCWB landscape. The optically driven acoustic impulses that consist of the superposition of the simultaneously excited $TR_{2m}$ ARs yield a set of impulsive qCWB fluctuations at specific temporal locations, which we could potentially adjust by changing the thickness of the AR-trapping optical fiber. The single-mode fiber cavity in which the cavity round-trip time is sufficiently longer than the AR decay time (~100 ns) and the optoacoustic effect is perturbative rather than dominant over the optical Kerr effect, is a unique, excellent platform for controlling and exploring a variety of MPD phenomena through optoacoustic manipulation of the qCWB. In this platform, in particular, we have experimentally shown that the optoacoustically induced impulsive qCWB fluctuations act as the point sources and attractive potentials for the bunched pulses and the SR. Hence, the optoacoustic qCWB fluctuations at the well-defined temporal positions offer a novel means to control the motion of a single drifting pulse or the emission and collision of SR, without complicated external injection of control signals into the laser cavity. Furthermore, such the controllability could be exploited to understand the SR dynamics further and unveil the mechanism of the SR generation that has not been clearly identified yet. Also, our observations provide new perspectives on studying ultrafast buildup of mode-locked pulses [43,44], as they suggest the boost of the mode-locked pulse formation from the continuous-wave beats mediated by the ARs. In addition, in sharp contrast to most of the previous works where only the $R_{0m}$ ARs were taken into account while the $TR_{2m}$ ARs were neglected, we experimentally verify that the $TR_{2m}$ ARs can play a critical role in mediating optoacoustically induced MPD phenomena. This might imply the existence of undiscovered nature of optoacoustic interactions in mode-locked fiber lasers. Our work opens up new opportunities to extend the field patterning [26,27] to the qCWB and the navigation to more intriguing MPD in the mode-locked fiber lasers.

## ACKNOWLEDGMENTS

This work was supported by the National Research Foundation of Korea (NRF) grants funded by the Korea government (MSIT) (NRF-2013R1A1A1007933, NRF-2016R1A2B4011862).

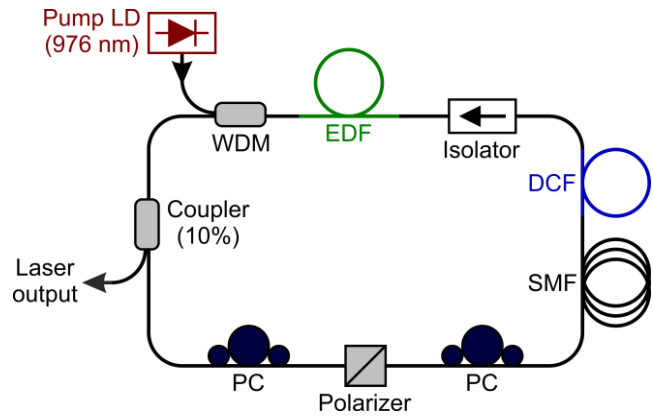

Fig. 1. Schematic diagram of our passively mode-locked fiber laser setup. LD, laser diode; WDM, wavelength division multiplexer; EDF, erbium-doped fiber; DCF, dispersion compensating fiber; SMF, single-mode fiber; PC, polarization controller.

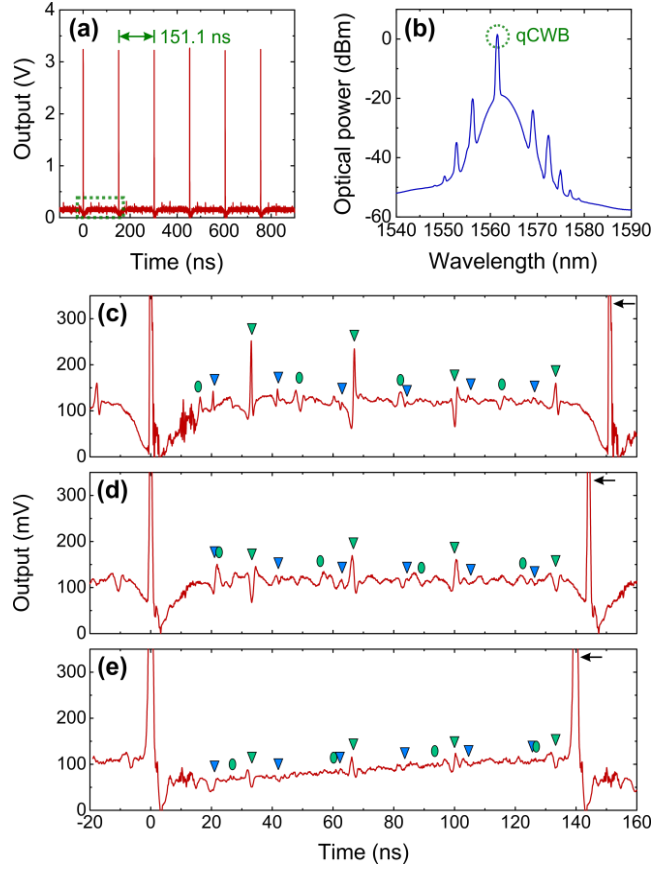

Fig. 2. Optoacoustically induced impulsive qCWB modulations. (a) Single-shot oscilloscope trace of the typical laser output that contains the mode-locked pulse train at the fundamental repetition rate (cavity round-trip time: $\tau_C$ = 151.1 ns) and the co-propagating qCWB. The pump power and the net cavity dispersion are set as $P_P$ = 300 mW and $\beta_C$ = -4.3 ps$^2$/km, respectively. (b) Typical optical spectrum of the laser output, where the spectral peak at 1561.5 nm corresponds to the qCWB, whereas the broad spectrum with the Kelly sidebands to the mode-locked pulses. (c) Zoomed-in oscilloscope trace that focuses on the qCWB (corresponding to the green dashed box in (a)), which is obtained with 1,000 times averaging for clear identification of the sharp intensity modulation of the qCWB. Green and blue triangles indicate the theoretically predicted temporal positions of integer multiples of $\tau_{TR}$ = 33 ns and $\tau_R$ = 21 ns, respectively, relative to the leftmost mode-locked pulse (at the time $t$ = 0). Green ellipses belong to the 33 ns branch of temporal locations from the mode-locked pulse in the previous round-trip (i.e., at $t$ = -151.1 ns). (d,e) qCWB fluctuations obtained in the same manner as (c), but at different values of $\tau_C$ and $\beta_C$ by changing the length of DCF. ($\tau_C$, $\beta_C$) = (144.2 ns, -8.7 ps$^2$/km) for (d) and ($\tau_C$, $\beta_C$) = (139.8 ns, -13.7 ps$^2$/km) for (e), while the pump power is fixed as $P_P$ = 300 mW. The black horizontal arrow in each plot of (c–e) stands for the main mode-locked pulse in the next round-trip.

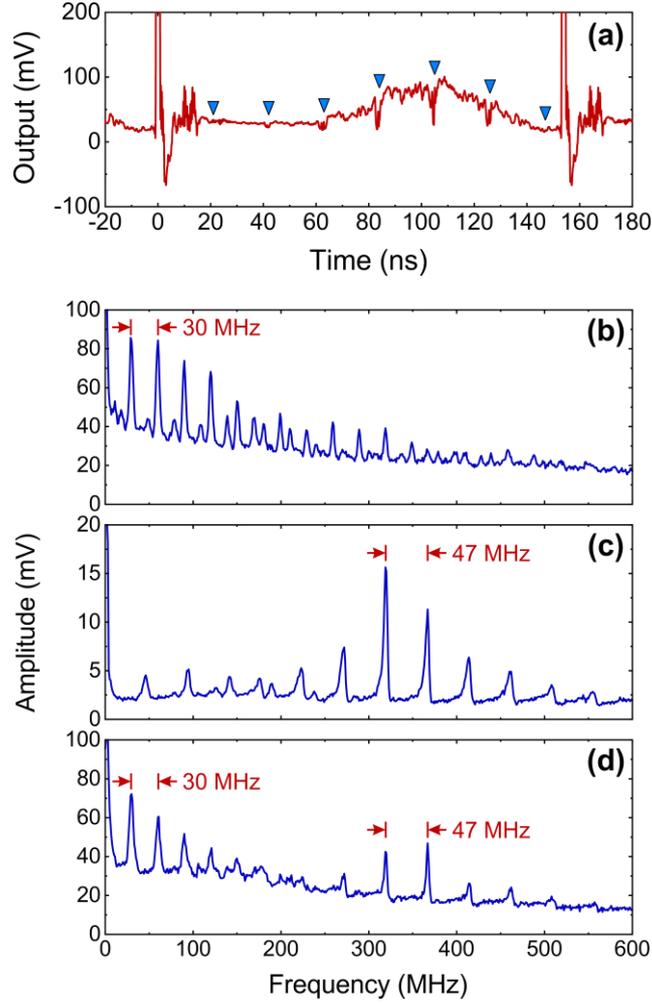

Fig. 3. Optoacoustically induced qCWB modulations from the laser excluding the intracavity polarizer. (a) Typical averaged oscilloscope trace that is zoomed-in around the impulsive qCWB modulations, as the laser is passively mode-locked by using a SESAM as a saturable absorber. The pump power is set as $P_P$ = 300 mW. Blue triangles indicate the theoretically predicted temporal locations of integer multiples of $\tau_R$ = 21 ns relative to the leftmost mode-locked pulse (at the time $t$ = 0). (b–d) Electrical spectra of the continuous-wave output from the laser in which passive mode locking is suppressed by removing both the intracavity polarizer and the SESAM. In (b), the spectrum exhibits several peaks at integer multiples of 30 MHz. In (c), the spectrum displays several peaks at integer multiples of 47 MHz. In (d), the spectrum contains both the 30-MHz-branch peaks and the 47-MHz-branch ones. The pump powers are $P_P$ = 80 mW for (c) and $P_P$ = 400 mW for (b) and (d). The resolution bandwidth of the RFSA is fixed as 2 MHz for (b–d).

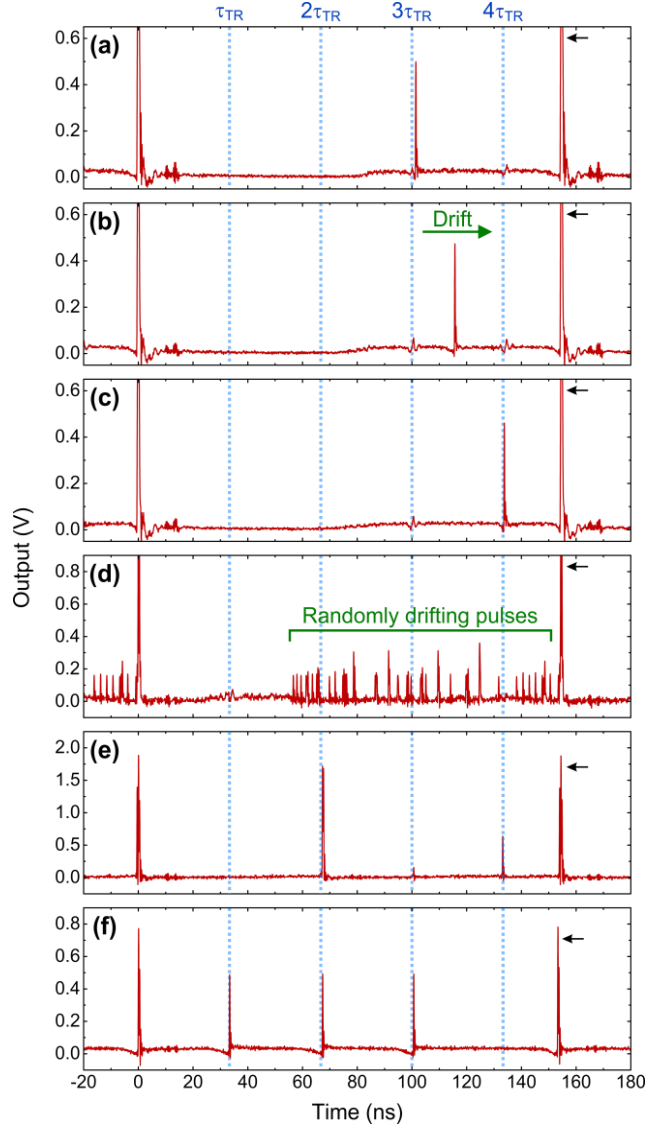

Fig. 4. Oscilloscope traces that display the dynamic multi-pulse phenomena involving a single drifting pulse bunch or a large set of randomly distributed ones on the qCWB landscape, which are strongly affected by the optoacoustic qCWB peaks. (a–c) Transport of a single pulse bunch between two neighboring qCWB peaks at $3\tau_{TR}$ = 100 ns and $4\tau_{TR}$ = 133 ns. The cavity parameters are $\tau_C$ = 154.6 ns, $\beta_C$ = -4.7 ps$^2$/km, and $P_P$ = 300 mW. In (a), a small pulse bunch is initially trapped at $3\tau_{TR}$ = 100 ns. In (b), when we perturb this state by slightly changing the intracavity PCs, the pulse bunch starts to drift toward the next main mode-locked pulse at $\tau_C$ = 154.6 ns (indicated by the black arrow). The drift speed is determined as ~0.5 m/s. Note the maintenance of both the qCWB peaks at $3\tau_{TR}$ = 100 ns and $4\tau_{TR}$ = 133 ns. In (c), the drifting pulse bunch is captured and stably held at the next optoacoustic qCWB peak at $4\tau_{TR}$ = 133 ns. (d) Typical multi-pulse state in which a large number of small pulses drift randomly over the qCWB landscape. (e,f) Two examples of final steady states of (d), where the set of small drifting pulses in (d) is organized eventually into multiple pulse bunches trapped at the optoacoustic qCWB peaks, which form regularly spaced pulse patterns with the spacing of $\tau_{TR}$ = 33 ns.

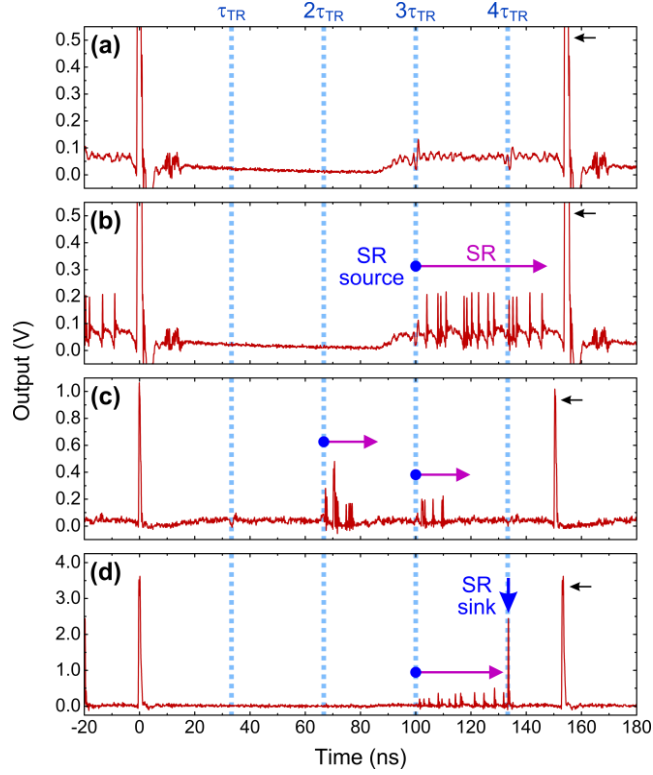

Fig. 5. Oscilloscope traces that show the novel SR states strongly affected by the optoacoustic qCWB peaks in diverse ways. (a) qCWB exhibiting optoacoustically induced sharp fluctuations at $3\tau_{TR} = 100$ ns and $4\tau_{TR} = 133$ ns. The cavity parameters are the same as in Fig. 4(a)–4(c), i.e., $\tau_C = 154.6$ ns and $\beta_C = -4.7$ ps$^2$/km, but the pump power is increased to $P_P = 570$ mW. (b) SR is emitted exclusively from a single optoacoustic qCWB peak at $3\tau_{TR} = 100$ ns, whereas the SR generation from the other arbitrary temporal positions is highly suppressed. This state is obtained by slightly changing the state of the intracavity PCs from (a). The SR collides eventually at the next main mode-locked pulse at $\tau_C = 154.6$ ns (indicated by the black arrow). (c) The SR can also be generated at multiple optoacoustic qCWB peaks at the same time. (d) When the drift speed of the SR is sufficiently low, it can collide at the next optoacoustic qCWB peak at $4\tau_{TR} = 133$ ns, rather than the main mode-locked pulse. The energy carried by the SR is then accumulated at the 'SR sink', which yields the growing pulse bunch trapped at the temporal position.

# Supplementary Information: Characterization of pure qCWB state

In addition to the qCWB co-circulating with mode-locked pulses, we also investigate the pure qCWB state, which we generate by suppressing the mode locking through the control of the intracavity polarization controllers. Here, we insert a 130-m-long strand of additional standard single-mode fiber in the laser cavity to enhance the optoacoustic interactions and reduce the cavity free spectral range, for more sensitive identification of the optoacoustic effect on the qCWB state. While the oscilloscope trace of the noisy qCWB state does not show any optoacoustic effect apparently (Fig. S1(a)), its electrical spectrum measured with a radio-frequency spectrum analyzer (RFSA) in the video averaging mode reveals several distinct spectral peaks (Fig. S1(b)). We summarize and compare the experimentally observed spectral peaks and the theoretically predicted AR frequencies [S1] in Table S1. It can be seen that the spectral peak positions determined from the electrical spectrum in Fig. S1(b) match far better the $TR_{2m}$ ARs than the $R_{0m}$ ones. There are also some unrevealed spectral peaks of the $TR_{2m}$ ARs for $m$ = 2, 4, 6, 14, and 17, which is attributed to their relatively small optoacoustic overlaps with the optical mode [S1,S2]. Although the $TR_{2m}$ ARs in Fig. S1 are thermally excited rather than electrostrictively driven by optical pulses, they are closely linked to the qCWB modulations in Fig. 2. The electrical spectrum in Fig. S1(b) represents a measure of the amount of polarization coupling that creates the qCWB modulations in Figs. 2(c)–2(e). The spectral peaks in Fig. S1(b) are spaced by 30 MHz, which corresponds to the inverse of the temporal spacing of the qCWB modulation of 33 ns. We note that the spectral measurement with an RFSA resolution higher than the cavity FSR (1.39 MHz in this case) reveals more spectral peaks equally spaced by the cavity FSR, where the locally highest peaks are placed close to the frequencies of the $TR_{2m}$ ARs, as shown in Figs. S1(c) and S1(d). It is noteworthy that the $TR_{2m}$ ARs can be categorized into two branches, the torsional and the radial ones. Between them, the torsional one has relatively larger torsional displacements that give rise to stronger polarization coupling. The frequencies of the $TR_{2m}$ ARs in this branch are separated by ~30 MHz, which is the inverse of the 33 ns duration that the torsional vibrational component takes to bounce back and forth within the 125-μm-thick silica-glass fiber cladding with the shear sound velocity of $V_S$ = 3,740 m/s [S3]. On the other hand, the $TR_{2m}$ ARs in the radial branch that exhibit relatively stronger radial displacements have the frequency spacing of ~47 MHz, which is the inverse of the 21 ns duration of the round-trip propagation along the 125-μm-thick cladding at the longitudinal sound velocity of $V_L$ = 5,996 m/s [S3].

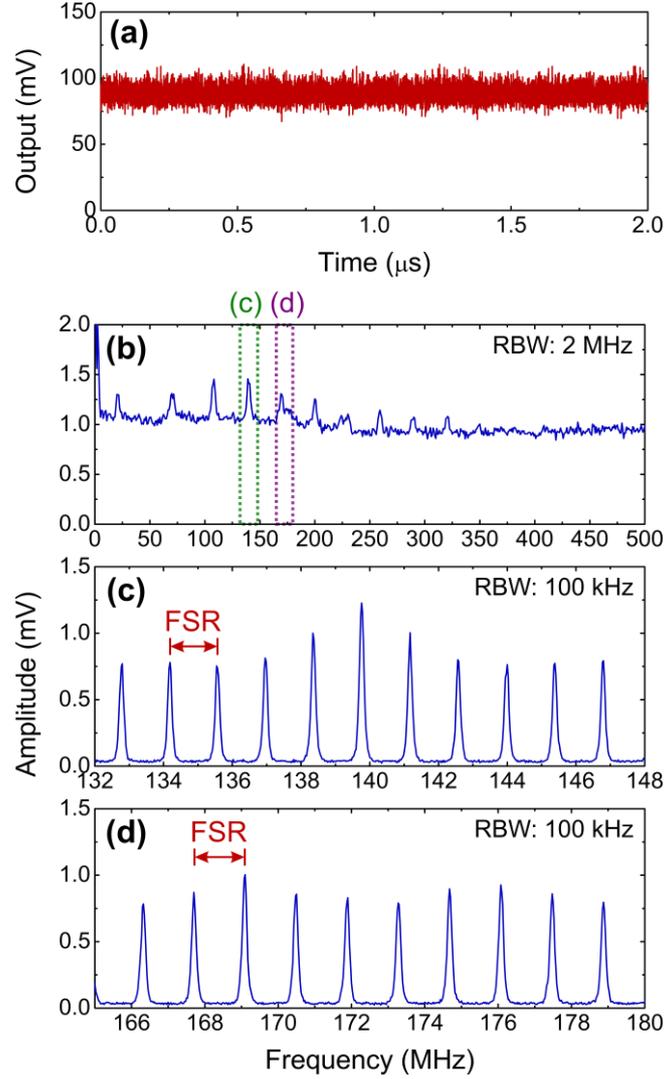

Fig. S1. Laser output in the pure qCWB state, the mode locking being suppressed via adjustment of the intracavity polarization controllers. Here, the cavity FSR is reduced to 1.39 MHz (corresponding to the total cavity length of 144 m) by inserting a 130-m-long strand of additional standard SMF in the laser cavity. (a) Oscilloscope trace of the laser output at the pump power of $P_P$ = 400 mW. (b) Electrical spectrum of the laser output in (a), measured with a radio-frequency spectrum analyzer (RFSA) in the video averaging mode with the resolution bandwidth of 2 MHz. See Table S1 for the comparison of the frequencies of the spectral peaks with the theoretically predicted ones. (c) Resolved electrical spectrum around the spectral peak at 139.76 MHz (indicated by the green dashed box in (b)), which reveals more spectral peaks equally spaced by the cavity FSR. (d) Resolved electrical spectrum around the spectral peaks at 169.10 MHz and 176.08 MHz (indicated by the purple dashed box in (b)). In (c) and (d), the locally highest peaks are nearby the AR frequencies. For these resolved measurements, the resolution bandwidth of the RFSA is enhanced as 100 kHz, while the video averaging mode is kept turned on.

Table S1. Comparison between the frequencies of experimentally observed spectral peaks in Fig. S1(b) and the theoretically predicted frequencies of the first 17 $TR_{2m}$ and $R_{0m}$ ARs.

| AR mode order, $m$ | Experimentally observed RF peaks (MHz) | Theoretical frequencies of the $TR_{2m}$ ARs (MHz) | Theoretical frequencies of the $R_{0m}$ ARs (MHz) |
|---|---|---|---|
| 1 | 20.99 | 22.32 | 30.52 |
| 2 | - | 39.46 | 82.06 |
| 3 | 71.28 | 70.72 | 130.74 |
| 4 | - | 81.65 | 179.03 |
| 5 | 107.64 | 108.45 | 227.18 |
| 6 | - | 126.81 | 275.26 |
| 7 | 139.76 | 139.97 | 323.31 |
| 8 | 169.10 | 169.29 | 371.34 |
| 9 | 176.08 | 176.91 | 419.36 |
| 10 | 199.79 | 200.19 | 467.36 |
| 11 | 224.91 | 224.68 | 515.36 |
| 12 | 230.47 | 230.77 | 563.35 |
| 13 | 259.85 | 260.33 | 611.34 |
| 14 | - | 273.59 | 659.33 |
| 15 | 289.23 | 290.55 | 707.31 |
| 16 | 319.97 | 319.88 | 755.29 |
| 17 | - | 322.51 | 803.27 |